\documentclass[amstex,useAMS,epsf,amssymb,usenatbib]{mn2e}
\usepackage{epsfig}
\usepackage{graphicx}
\usepackage{color}
\usepackage{fmtcount}
\usepackage{amssymb,amsmath}
\usepackage{comment}
\usepackage{multicol}
\usepackage{soul}
\usepackage{xcolor}
\sethlcolor{yellow}

\title[Simulating SZ intensity maps]{Simulating SZ intensity maps of giant AGN cocoons}

\author[D. A. Prokhorov et al.]{D. A. Prokhorov$^{1}$\thanks{E-mail:
phdmitry@stanford.edu}, A. Moraghan$^2$\thanks{E-mail:
moraghan@galaxy.yonsei.ac.kr},  V. Antonuccio-Delogu$^{3,4}$ , and J. Silk$^{5,6,7}$
\\
~\\
$^{1}$W. W. Hansen Experimental Physics Laboratory, Stanford
University, Stanford, CA 94305, USA\\
$^{2}$ Center for Galaxy Evolution Research and Department of Astronomy, Yonsei University, Seoul 120-749, Republic of Korea\\
$^{3}$ INAF – Osservatorio Astrofisico di Catania, via S. Sofia 78, 95123 Catania, Italy
\\
$^{4}$ Scuola Superiore di Catania, via San Nullo 5/i, 95123 Catania, Italy\\
$^{5}$ Astrophysics, Department of Physics, University of Oxford, Keble Road Ox1 3RH, Oxford, UK\\
$^{6}$ Institut d’Astrophysique de Paris, 98bis Bd Arago, 75014 Paris, France\\
$^{7}$ Department of Physics and Astronomy, The Johns Hopkins University, 2700 San Martin Drive, Baltimore MD 21218, USA\\}

\date{Accepted .....
      Received ..... ;
      in original form .....}

\pagerange{\pageref{firstpage}--\pageref{lastpage}}
\pubyear{2012}

\begin{document}

\maketitle

\label{firstpage}

%%%%%%%%%%%%%%%%%%%%%%%%%%%%%%%%%%%%%%%%%%%%%%%%%%%%%%%%%%
\begin{abstract}
We perform relativistic hydrodynamic simulations of the formation and evolution of AGN cocoons produced by
very light powerful jets. We calculate the intensity maps of the Sunyaev--Zel'dovich (SZ) effect at high frequencies 
for the simulated AGN cocoons using the relativistically correct Wright formalism. Our fully relativistic calculations 
{demonstrate} that the contribution from the high temperature gas ($k_{\rm{b}} T_{\rm{e}}\simeq 100$ keV) to the SZ signal from AGN cocoons at high frequencies is stronger than that  from the shocked ambient intercluster medium 
{owing to the fact that the relativistic spectral functions  peak at these temperature values}.
We present  simulations of the SZ effect from AGN cocoons at various frequencies, and demonstrate that SZ observations at 217 GHz and at higher frequencies, such as 857 GHz, will provide us with
knowledge about the dynamically-dominant component of AGN cocoons.

\end{abstract}
%%%%%%%%%%%%%%%%%%%%%%%%%%%%%%%%%%%%%%%%%%%%%%%%%%%%%%%%%%

%%%%%%%%%%%%%%%%%%%%%%%%%%%%%%%%%%%%%%%%%%%%%%%%%%%%%%%

\section{Introduction}

The standard evolutionary scenario for AGNs suggests that their jets are not in direct  contact with the intergalactic medium (IGM), but rather are enveloped in a cocoon. 
The cocoon around a pair of supersonic, low-density (when compared to the ambient IGM) 
jets acts as a `wastebasket' for most of the energy deposited by the jets (Scheuer 1974). 
{A number of cavities in X-ray surface brightness maps were detected by the} {\it{Chandra}} {X-ray observatory in clusters of galaxies (for a review, see McNamara \& Nulsen 2007).}
{So far the largest} cavities in X-ray surface brightness maps {have been revealed} in the MS 0735+7421 and Hercules A clusters of galaxies (McNamara et al. 2005; Nulsen et al. 2005). These X-ray cavities extend over several hundreds of 
kpc and are formed by powerful AGN jet activity. The presence of X-ray cavities
can be caused by  very hot gas embedded in the AGN cocoons. 
Low-density gas with high temperatures of  $k_{\mathrm{b}} T_{\mathrm{e}}\simeq 100$ keV does not significantly 
emit in the soft X-ray band and, therefore, high temperature regions can be associated with 
X--ray cavities. Numerical simulations of AGN cocoons in galaxy clusters show that electron 
temperatures  of the order of 100 keV are expected for the plasma in AGN cocoons (see, e.g., 
Sternberg \& Soker 2009; Perucho et al. 2011).   

Inverse Compton (IC) scattering of cosmic microwave background (CMB) photons
by free thermal electrons located in an AGN cocoon causes a change of the CMB spectrum 
(see Pfrommer et al. 2005 and Prokhorov et al. 2010). In this paper,
we identify this change of the CMB spectrum with the Sunyaev--Zel'dovich (SZ) effect 
{(Zel'dovich \& Sunyaev 1969; for a review, see Birkinshaw 1999)} from AGN cocoons.
The advent of new telescopes, such as {\it{Herschel}}, {\it{GBT}}, and {\it{ALMA}} should allow us to measure the 
CMB distortion towards AGN cocoons at various frequencies with high sensitivity 
and with high angular resolution (see, Zemcov et al. 2010; Korngut et al. 2011; Yamada et al. 2012, 
for the first observations of the SZ effect from galaxy clusters by {\it{Herschel-SPIRE}}, {\it{GBT-MUSTANG}}, and 
the recent simulations of SZ maps of {\it{ALMA}}, respectively). As shown by Prokhorov et al. (2010), the CMB intensity 
change at a frequency of 217 GHz, which approximately corresponds to the crossover frequency of the 
SZ effect from a low temperature gas with $k_{\mathrm{b}} T_{\mathrm{e}} < 5$ keV (see Sunyaev \& Zel'dovich 1980), 
is maximal at the temperature of $\simeq$100 keV (see Colafrancesco 2005 for a review of the SZ effect from AGN cocoons 
in non-thermal electron models). The frequency shift of CMB photons owing to IC scattering increases with 
electron temperature.    
Therefore, we assume that the CMB distortion caused by  IC scattering by a high temperature 
plasma with $k_{\mathrm{b}} T_{\mathrm{e}}\simeq 100$ keV should also be significant at high 
frequencies.  

So far, the SZ effect from AGN cocoons in galaxy clusters has been calculated by using   
{analytical toy models} for gas pressure and temperature distributions in cocoons 
(see Colafrancesco 2005; Pfrommer et al. 2005) because of lack of available
hydrodynamic simulations of AGN cocoons. {These toy models do not take into account intrinsic properties of the jet--ICM system and, therefore, can lead to oversimplification. To test 
consistency and feasibility of these toy models, it is necessary to calculate the SZ effect from AGN cocoons which are derived from relativistic hydrodynamic simulations of the jet--ICM system and
to confront the results obtained from the analytical toy models with those obtained from more realistic numerical models.}
Using the \textsc{PLUTO} code (Mignone et al. 2007), we performed relativistic hydrodynamic simulations of 
the formation and evolution of AGN cocoons produced by very light {(i.e. under-dense
compared with the ICM)} powerful jets. 
Our numerical simulations take into account different properties of AGN jets and ICM, 
such as the power of the jets, {the jet-to-ICM density contrast}, velocity, gas density and temperature distributions, and a dark matter halo profile, all providing us with a more realistic AGN cocoon model. {The performed simulations demonstrate that the
internal structure of the region around AGN jets is more complex than that assumed in analytical toy models.}
We calculate the CMB spectrum distortion caused by IC scattering of the CMB photons by highly energetic electrons 
in the framework of the relativistically correct Wright formalism and study the SZ effect from simulated AGN cocoons
at various frequencies. 
In this paper, we show that more realistic plasma models obtained from numerical hydrodynamic 
simulations {predict: (1) that the morphology of SZ intensity maps changes with frequency and (2) that SZ signals at high frequencies from AGN cocoons are higher 
than those from the shocked ambient intracluster medium and, therefore, the SZ morphological study can reveal the presence of AGN cocoons on SZ maps}. The fully relativistic approach for calculating 
the SZ effect from AGN cocoons considered below shows that the cocoon is strongly inhomogeneous
and that future measurements of the SZ effect will permit one to study the internal structure of regions formed by AGN activity. {We also show that the tight constraints on the total thermal energy of high temperature electrons which are stored in AGN cocoons can be obtained by taking into account the shape of relativistic SZ spectral functions.}
 
The layout of the paper is as follows. We describe the relativistic hydrodynamic simulations
in Sect. 2. We calculate the SZ effect from the simulated AGN cocoons in the relativistically correct Wright formalism {and study the possibility to detect AGN cocoons through SZ observations} in Sect. 3. Our conclusions are presented in Sect. 4.

%%%%%%%%%%%%%%%%%%%%%%%%%%%%%%%%%%%%%%%%%%%%%%%%%%%%%%%
\section{Relativistic hydrodynamic simulations of AGN cocoons}
%%%%%%%%%%%%%%%%%%%%%%%%%%%%%%%%%%%%%%%%%%%%%%%%%%%%%%%
\label{simulations}

In this section, we introduce the code and initial conditions used in our simulations, 
describe the setup of the simulated domain, ambient medium and jet parameters, 
and present our results of the relativistic hydrodynamic simulations of AGN cocoons.

\subsection{Simulation setup}
%\textbf{Relativistic hydrodynamics with the PLUTO code.} 
Our numerical simulations of AGN outflows were performed using the publicly available
\textsc{PLUTO} astrophysical computational code (Mignone et al. 2007).
\textsc{PLUTO} is a modern multi-physics, high-resolution, high-order shock capturing Godunov code with
the later versions including Adaptive Mesh Refinement (AMR) capability (Mignone et al. 2012).
It was developed with a modular design and includes a variety of physics and solver
algorithms making it very adaptable to tackle a diverse range of astrophysical problems (see,
e.g. Murphy et al. 2010; Kritsuk et al. 2011; Te{\c s}ileanu et al. 2012.)
It is particularly suitable for our specific purposes of AGN jet simulations 
as it incorporates a module to solve the equations of special Relativistic 
Hydrodynamics (RHD), which we utilise to simulate our relativistic outflows.

As we are only interested in the final large-scale evolved morphology of the AGN jet/cocoon system, we assume the launching, acceleration, and collimation process has taken place off the grid and so we only introduce a pre-collimated jet flow into the computational domain.
Therefore, we use the RHD module where we also assume that the bulk kinetic energy within the flow has overcome any possible MHD effects on the cocoon morphology. {Magnetic pressures in the ICM are likely to be much smaller than the gas pressure, but magnetic fields can still play a significant role on scales (e.g., Jones 2008 and references therein) which are much smaller than our computational domain}.

The \textsc{PLUTO} RHD module solves the equations of relativistic hydrodynamics as described in Mignone et al. (2007). 
The equations are closed with the \textsc{PLUTO} `TAUB' equation of state, which is the quadratic approximation of the
theoretical relativistic perfect gas equation of state.
%The equations are closed by the `TM' Equation of State for relativistic gas which was proposed by Mignone et al. (2005).
In this work we did not deem it necessary to include a radiative cooling function as the typical
cooling times of the low density, high temperature gas present in AGN cocoons are significantly
longer compared to the simulation time and should not substantially modify the result of our SZ effect study 
(see Appendix B by Antonuccio-Delogu \& Silk 2008). 
%but will do so in a future publication utilising a more advanced physical model.
We allow the code to solve the RHD equations using the \textsc{MUSCL}-Hancock second-order scheme with
the \textsc{HLLC} Riemann solver by Mignone \& Bodo (2005), {and using the
second-order linear Total Variation Diminishing (TVD) interpolation}.
This combination promises the most efficient second-order scheme, plus a robust and accurate Riemann solver.
We make use of the common approximation that astrophysical jets can be treated as cylindrical
and thus we perform our simulations using 2.5D axisymmetry.
We also include AMR which adaptively applies a higher resolution grid only where required which can be a much 
more efficient alternative to using a fixed grid of the highest resolution over the entire computational domain.

\begin{table}
\begin{center}
\caption{Model parameters}
\label{model_parameters}
\begin{tabular}{@{}l c c c c}
\hline
Parameter names & Model A & Model B\\
\hline
Density contrast (n$_{\rm{jet}}$/n$_{\rm{ICM}}$) & $5\times 10^{-5}$ & $5\times 10^{-5}$\\
Jet velocity & 0.99c & 0.985c\\
Jet power (erg s$^{-1}$) & $7.6\times10^{46}$  & $3.8\times10^{46}$ \\
Duration of jet's active phase & 25 Myr & 50 Myr\\
Time since end of active phase & 55 Myr & 50 Myr\\
\hline
\end{tabular}
\end{center}
\end{table}

\begin{itemize} \item \textbf{Initial Conditions.}
We have constrained our initial conditions based on observations of the AGN outburst in the
MS 0735+7421 galaxy cluster as described by Gitti et al. (2007) who
analysed observations with the XMM-Newton satellite.
The MS 0735+7421 outburst, with a red-shift of 0.22, is one of the most energetic AGN outflows observed
containing an estimated 1.2$\times$10$^{62}$erg of energy,
a dynamical lifetime of 100 Myr, and covering a 700 kpc diameter about the cluster (Gitti et al. 2007).
The high-power, large extent, and relative vicinity {(providing a scale of $1^{\prime}=213$ kpc at the cluster red-shift)} make this a suitable target for high-resolution SZ effect
observations of AGN cocoons.

A collimated jet beam with a zero degree opening angle is injected into the axisymmetric computational domain
within the region $r < 1$ at $z = 0$, where $r$ is the radial distance (in kpc) and $z$ is the distance along the axis of
symmetry.
Reflecting boundary conditions are imposed along the jet axis ($z$ axis),
and outside the jet radius for $r > 1$ at $z = 0$.
The remaining two boundaries possess outflow conditions.
The physical jet beam radius is set as 1 kpc and the axisymmetric computational domain covers a
physical range of 200$\times$400 kpc.
This scale is large enough to contain one-half of the MS 0735+7421 bi-polar outflow.
In our AMR simulations we set the coarse level 0 grid as 200$\times$400 zones and allow for three levels of refinement 
which provides our simulations with an effective resolution of 1600$\times$3200 zones with 125 pc per zone and 8 zones 
per jet radius. This resolution is sufficient to study the spatial distribution of thermodynamic
variables (e.g. pressure and temperature) in AGN cocoons.\\

\begin{figure}
\centering
\includegraphics[angle=0, width=9.0cm, height=6.0cm]{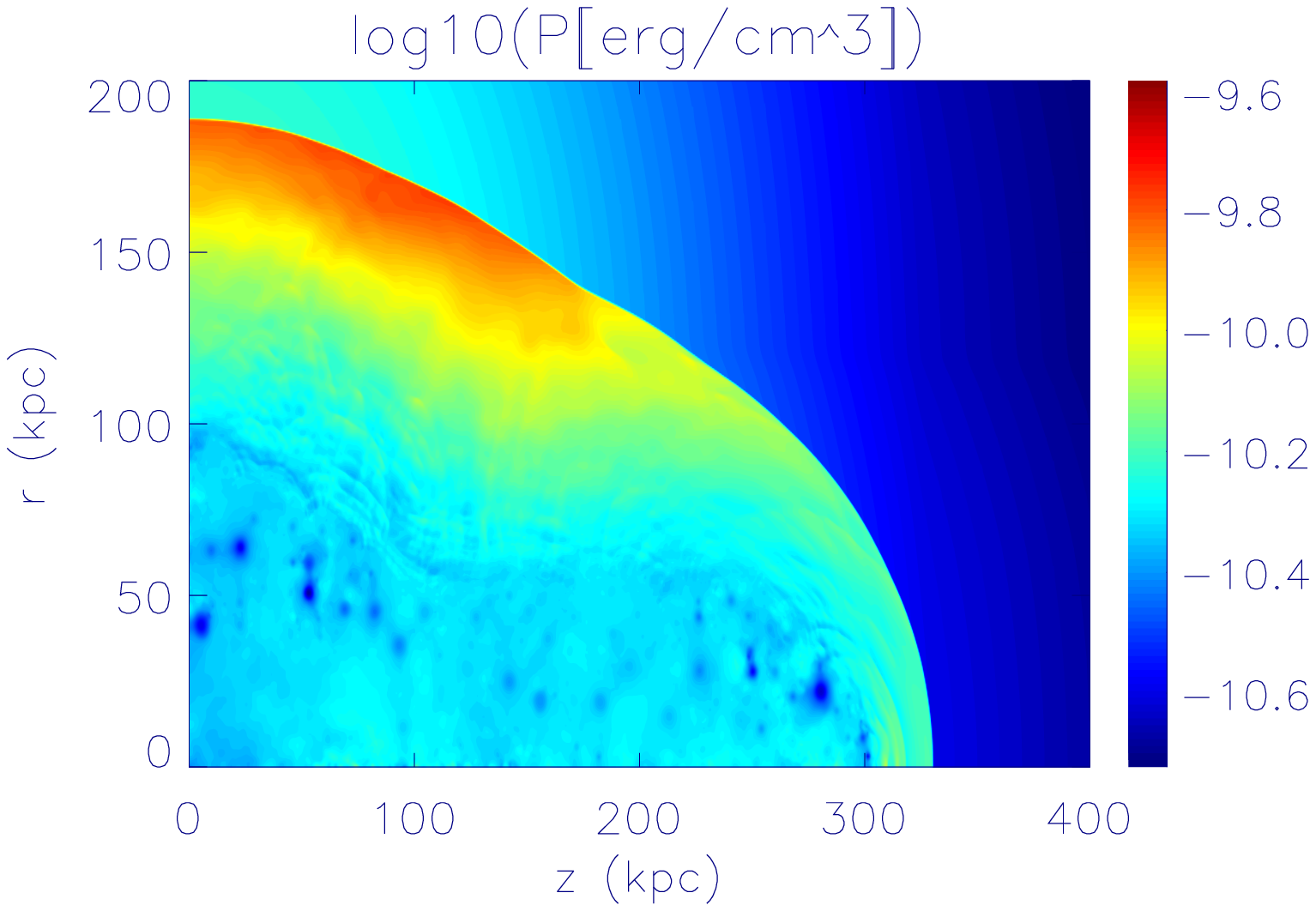}
\includegraphics[angle=0, width=8.5cm]{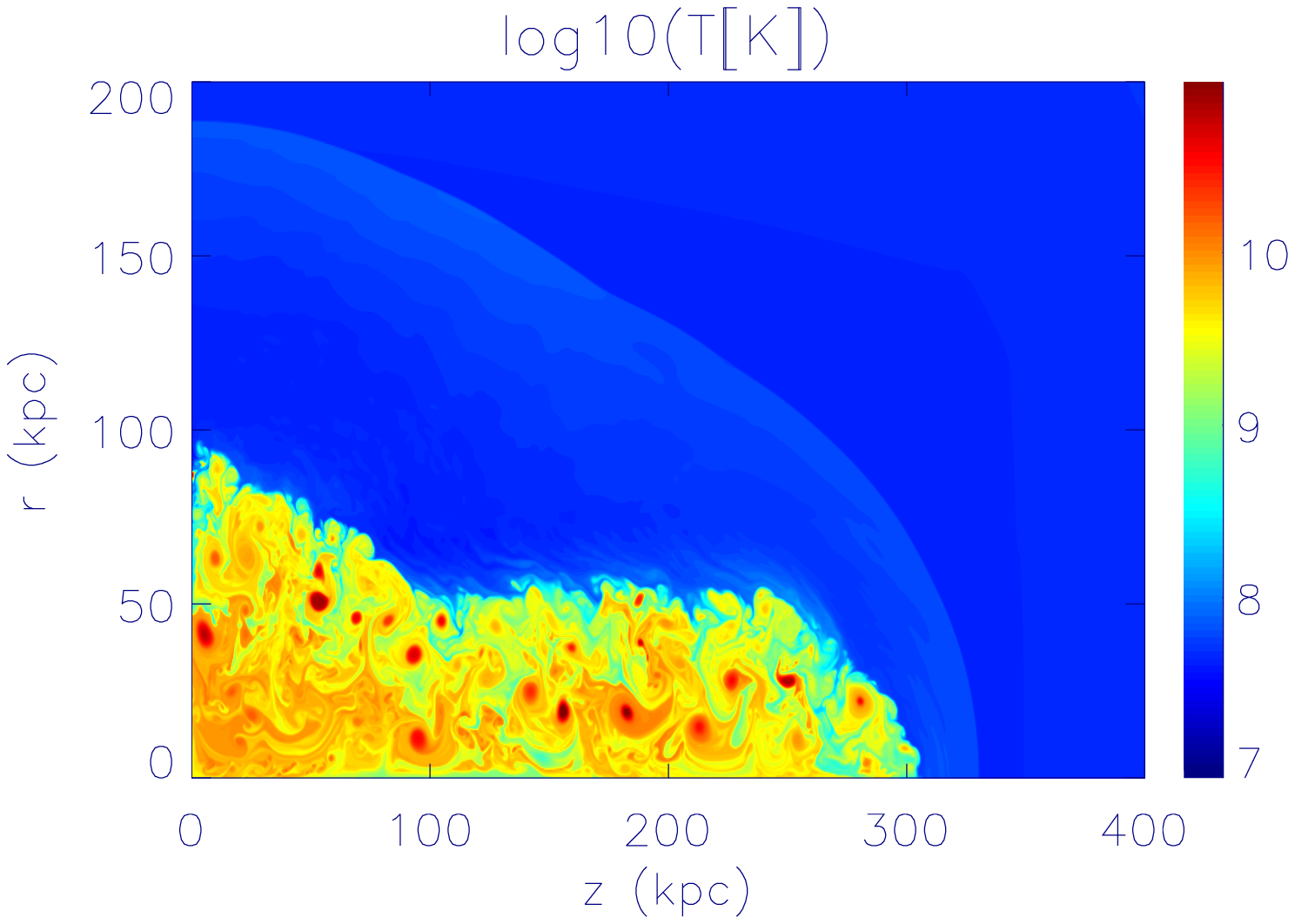}
\caption{The simulation map of the electron log-pressure (upper panel) and log-temperature (lower panel) 
for Model A.} \label{PT1}
\end{figure}

\begin{figure}
\centering
\includegraphics[angle=0, width=9.0cm, height=6.0cm]{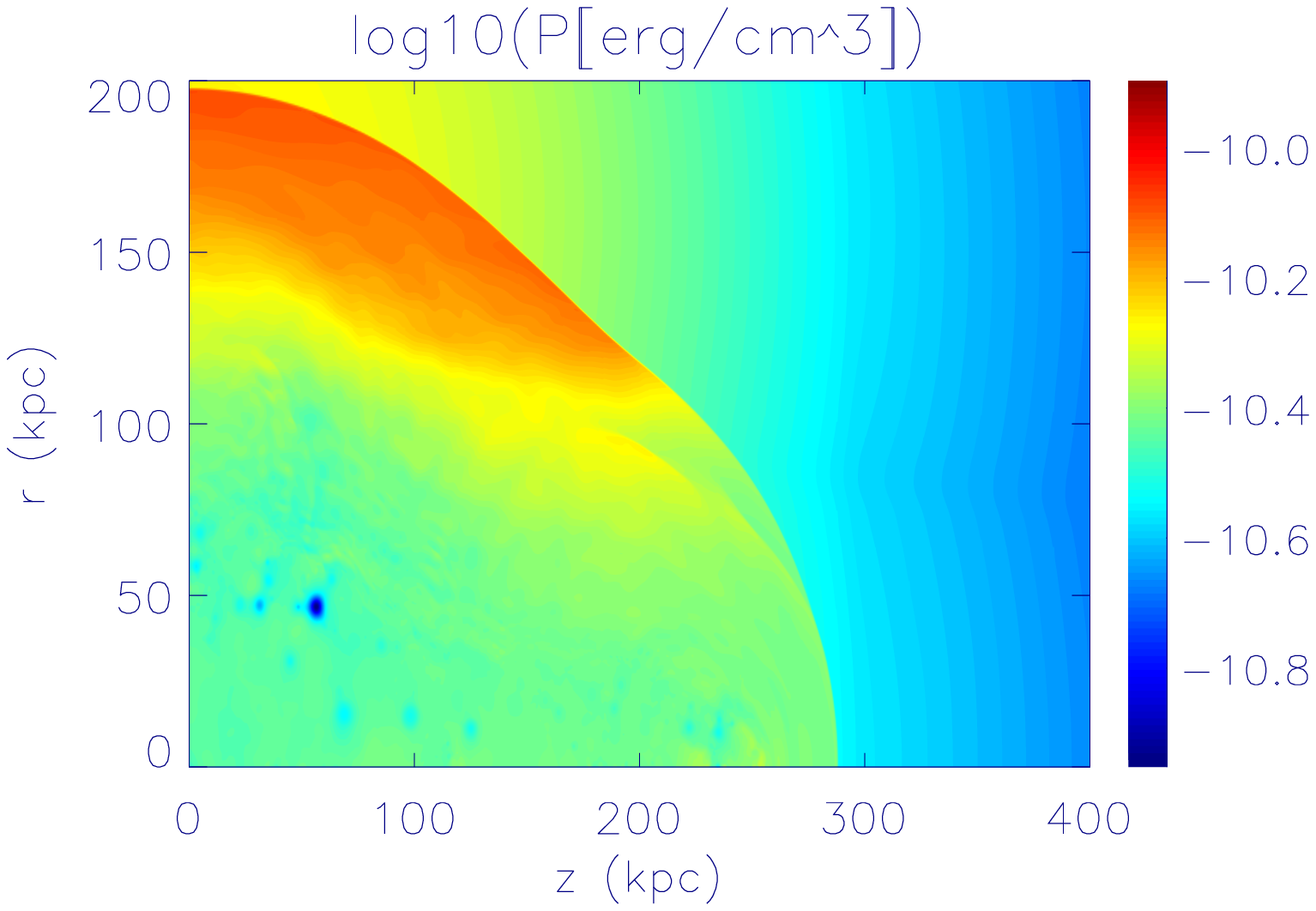}
\includegraphics[angle=0, width=8.5cm]{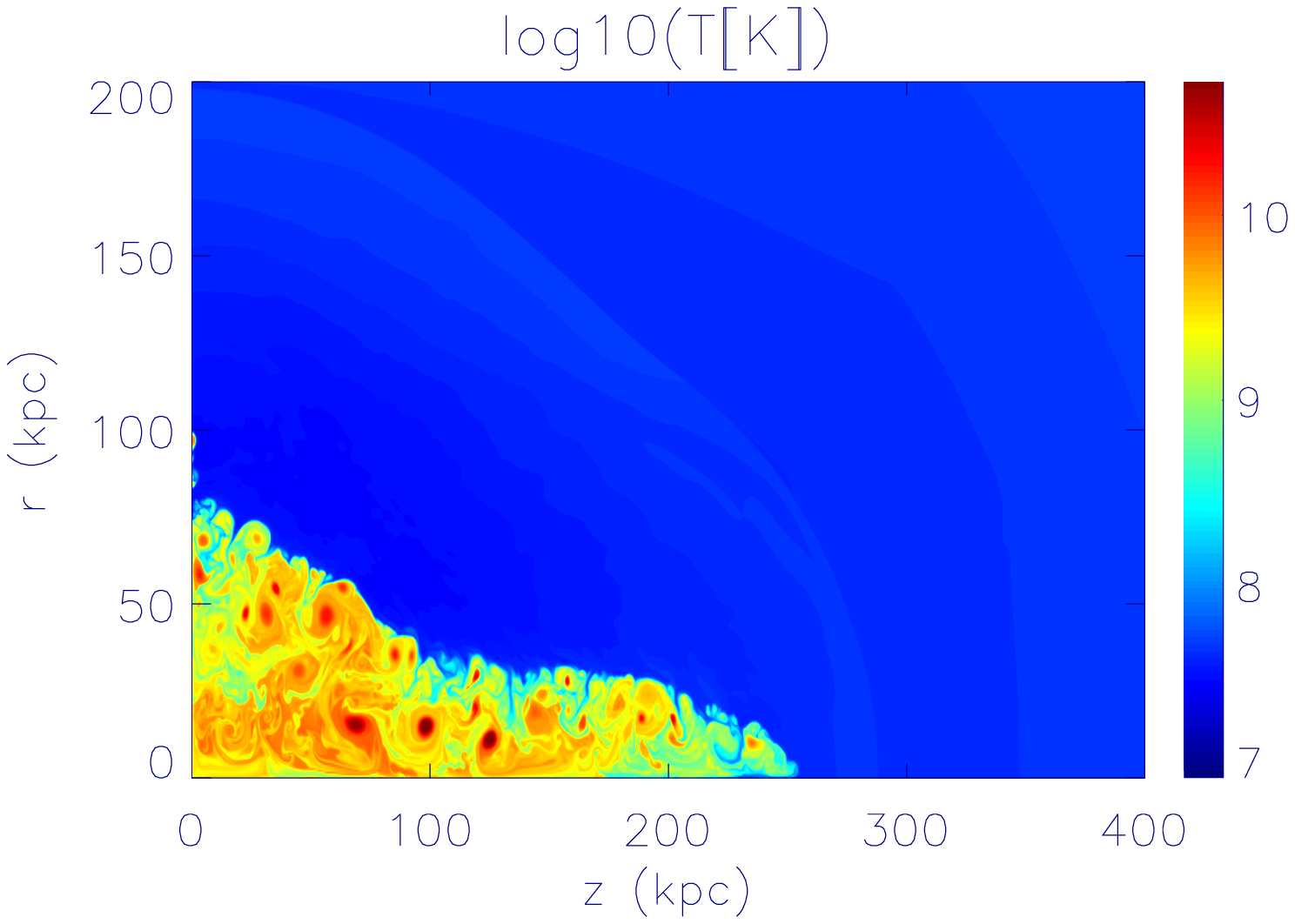}
\caption{The simulation map of the electron log-pressure (upper panel) and log-temperature (lower panel) 
for Model B.} \label{PT2}
\end{figure}

To add to the physical accuracy of our simulations, we include an ambient medium density and pressure profile
as determined from the electron-density surface brightness measurements and pressures of Gitti et al. (2007).
The authors used an analytical $\beta-$model expression (Cavaliere \& Fusco-Femiano 1976) to fit the observed density profiles.
We take the values of core radius, $r_c$, estimated as 195 kpc,
the gas density at the core, $n_{0,gas}$, measured to be 0.02 particles per cm$^3$,
and the beta-model fitting parameter $\beta$ determined to be 0.77 for a best
fit to the observational data (Gitti et al. 2007).
We also include the effects of a surrounding dark matter halo by adding a time-constant gravitational 
potential to the code.
Our gravitational potential, is based on the Navarro, Frenk, \& White (1996) (NFW) profile 
with the scale radius of the NFW profile, $r_s$, set as 498 kpc, 
and a concentration parameter of 3.45 derived by Gitti et al. (2007).
A similar set of our initial {ambient} conditions were used in the simulations of Sternberg \& Soker (2009).
The initial distributions of gas pressure and temperature in this region is not very certain, 
because AGN activity disturbs the ICM and because X-ray measurements only permit us to study the present
gas pressure and temperature distributions. However, we assume the measured pressure and temperature radial 
profiles from Gitti et al. (2007) are a good first approximation due to lack of more certain 
information about the initial gas distribution.

\item \textbf{Jet beam parameters.}
The precise composition of AGN jets is still uncertain although it is accepted that they are 
under-dense compared to the ambient medium.
Our aim is to inject a total energy of 6$\times 10^{61}$ erg onto the grid, 
which equals half of the energy in the observed MS0735+7421 system, and to obtain a crossing distance 
no greater than 400 kpc within the estimated dynamical lifetime of $\sim$100Myr.
The jet beam density and power, and hence velocity, are free parameters.
We thus present two models with different jet lifetimes or `duty cycles';
Model A where the jet is active for 25 Myr, and Model B where the jet is active for 50 Myr.
Both models possess an under-dense baryonic jet beam with a density of $10^{-6}$ particles per cm$^3$.
Using an expression relating the kinetic jet power and jet velocity within the relativistic 
regime, the 25 Myr jet of Model A injects a power of $7.6\times10^{46}$ erg s$^{-1}$ and has a velocity of 0.99c.
In contrast, the longer lifetime jet of the 50 Myr Model B injects a power of $3.8\times10^{46}$ erg s$^{-1}$ 
leading to a lower velocity of 0.985c. After the jet's active phase ceases, the simulation continues to 
evolve until the leading bowshock reaches a grid boundary.
A list of the relevant parameters of the simulations are described in Table \ref{model_parameters}.
We are following the traditional approach of highly relativistic jets (see also Perucho et al. 2011).
It should be noted that Sternberg \& Soker (2009) attempted to explain the
MS 0735+7421 outburst morphology using massive slow jets with wide opening angles which inflate
a more spherical X-ray cavity. 
\end{itemize}

\subsection{Simulated pressure and temperature maps.}
We plot the simulated electron pressure and temperature maps in Figs. \ref{PT1} and \ref{PT2} for Models A and
B, respectively. The upper panels of Figs. \ref{PT1} and \ref{PT2} show that the {contact} discontinuity
separates the region highly disturbed by AGN jet activity from the outer ICM.
Figs. \ref{PT1} and \ref{PT2} also show that the pressure values are higher in the outer layers of the
disturbed region and that the temperature values are higher in the inner layers of this region. 
Thus, we can distinguish an internal low pressure, a very high temperature cocoon and 
a shocked ambient gas region, which is externally bounded by a discontinuity with the outer ICM.
These regions are also dynamically very different. The region containing the shocked gas has on average
an expansion motion, while in the cocoon a large-scale circulation parallel and opposite to the main stream 
of the jet develops, originating from gas reflected away from the region near the hotspot. 
The turbulence within the cocoon surrounding the propagating jet  is generated naturally by the 
interaction of the jet with the ICM (see, also Antonuccio-Delogu \& Silk 2008). 
One can conclude from Figs. \ref{PT1} and \ref{PT2} that the density within the cocoon is very low compared
to that in the shocked ambient gas region,  because the temperatures are on average significantly higher
in the simulated cocoons. We checked and found that the radiative cooling time exceeds the simulation
time by at least one order of magnitude everywhere in the computational domain and, therefore, radiative
cooling has no significant effect on the evolution of the simulated AGN cocoons.

{The simulations show that the gas pressure in a very high temperature cocoon is
lower than that in a shocked ambient gas region and, therefore, 
we conclude that the assumption of pressure equilibrium in the jet--ICM system 
made by Pfrommer et al. (2005) leads to overestimation of the SZ effect from an AGN cocoon.
This is owing to the fact that the SZ effect is proportional to gas pressure integrated
along the line-of-sight.}

The temperature values in the simulated cocoons are $\sim 10^9-10^{10}$ K and are
in  agreement with those derived in  previous simulations by Antonuccio-Delogu \& Silk (2008),
Sternberg \& Soker (2009), and Perucho et al. (2011). 
{Note that the pressure and temperature values in AGN cocoons obtained
from our simulations (that assume very light jets) are similar to those obtained from the
simulations (that assume massive jets) by Sternberg \& Soker (2009). }
From the upper and lower panels of Figs. \ref{PT1} and \ref{PT2}, we find that the
temperature is changing within three orders of magnitudes over the whole simulated
region, whereas the pressure is only changing within one order of magnitude. This
is in agreement with the results of Krause (2003), where the detailed analysis
of pressure--density histograms calculated for the simulated cocoons was presented
and shows that very light jets provide their own pressure everywhere in a cocoon.
In Sect. 3, we will show that this fact is important to estimate the SZ signal from
simulated AGN cocoons.

\section{SZ effect from the simulated giant AGN cocoons}

In this section, we briefly describe both the non-relativistic and relativistically correct
formalisms to calculate the SZ effect emphasising a significant contribution to the SZ signal at high
frequencies from electrons with high temperatures. We calculate the SZ effect from the simulated AGN cocoons
in the framework of the relativistic Wright formalism. We present the simulated SZ maps of AGN cocoons
at various frequencies. The first relativistic simulations of the SZ effect from AGN cocoons 
form  the main results of our paper. 

%%%%%%%%%%%%%%%%%%%%%%%%%%%%%%%%%%%%%%%%%%%%%%%%%%%%%%%%%%%%%%%
\subsection{Comptonization of the CMB at high frequencies by high energy electrons}
%%%%%%%%%%%%%%%%%%%%%%%%%%%%%%%%%%%%%%%%%%%%%%%%%%%%%%%%%%%%%%%
 
High energy electron populations cause spectral CMB distortions through IC scattering which 
depend on electron temperature. Below we briefly describe the spectral CMB distortions that are
caused by non-relativistic electrons with $k_{\mathrm{b}} T_{\mathrm{e}}\lesssim 10$ keV 
(this case is known as the `standard' thermal SZ effect) and those are caused by relativistic electrons with 
$k_{\mathrm{b}} T_{\mathrm{e}}\gg  10$ keV. Particularly, we study the SZ effect at high frequencies
produced by high energy populations of electrons with $k_{\mathrm{b}} T_{\mathrm{e}}\simeq  100$ keV.

\begin{figure}
\centering
\includegraphics[angle=0, width=8.0cm]{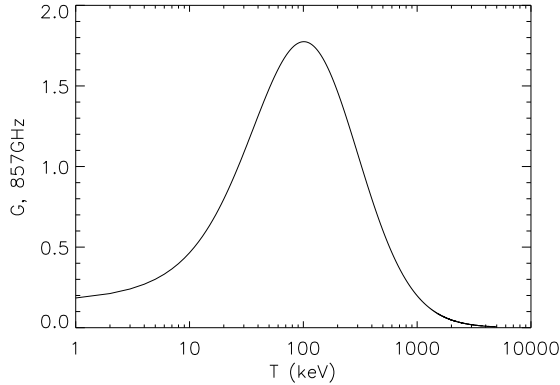}
\caption{Dependence of the relativistic spectral function at the frequency of 857 GHz on electron temperature.
} \label{F1}
\end{figure}

The CMB spectral distortion caused by IC scattering of CMB photons by electrons 
with temperatures of $k_{\mathrm{b}} T_{\mathrm{e}}\lesssim 10$ keV (the SZ effect) 
is described by the Kompaneets equation (Kompaneets 1957; Sunyaev \& Zel'dovich 1980) and is given by
\begin{equation}
\Delta I(x)=I_{0} \frac{\sigma_{\mathrm{T}}}{m_{\mathrm{e}}c^2} g(x) \int n_{\mathrm{e}} k_{\mathrm{b}}T_{\mathrm{e}}dl 
\end{equation} with the spectral dependence of 
\begin{equation}
g(x)=\frac{x^4 \exp(x)}{(\exp(x)-1)^2}\left(x\frac{\exp(x)+1}{\exp(x)-1}-4\right)
\end{equation}
where $x=h\nu/k_{\mathrm{b}}T_{\mathrm{cmb}}$ and $I_{0}=2(k_{\mathrm{b}} T_{\mathrm{cmb}})^3/(hc)^2$,
the integral is taken along the line-of-sight, $T_{\mathrm{e}}$ is the electron temperature, $n_{\mathrm{e}}$ 
is the electron number density, $\sigma_{\mathrm{T}}$ is the Thomson cross-section, $m_{\mathrm{e}}$ the electron mass, 
$c$ the speed of light, $T_\mathrm{cmb}=2.725$ K and $h$ the Planck constant. 

\begin{figure*}
\centering
\includegraphics[angle=0, width=8.5cm]{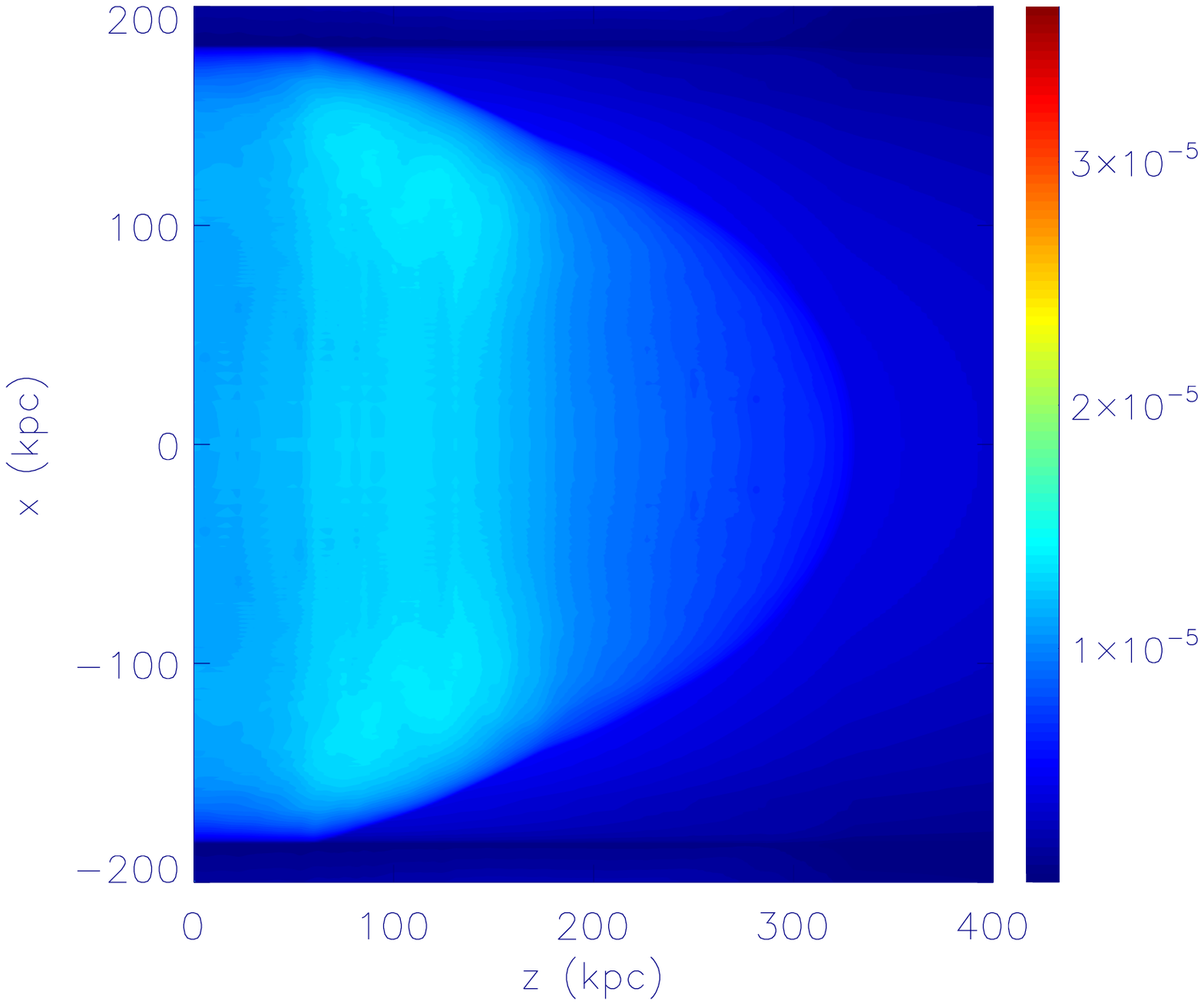}
\includegraphics[angle=0, width=8.5cm]{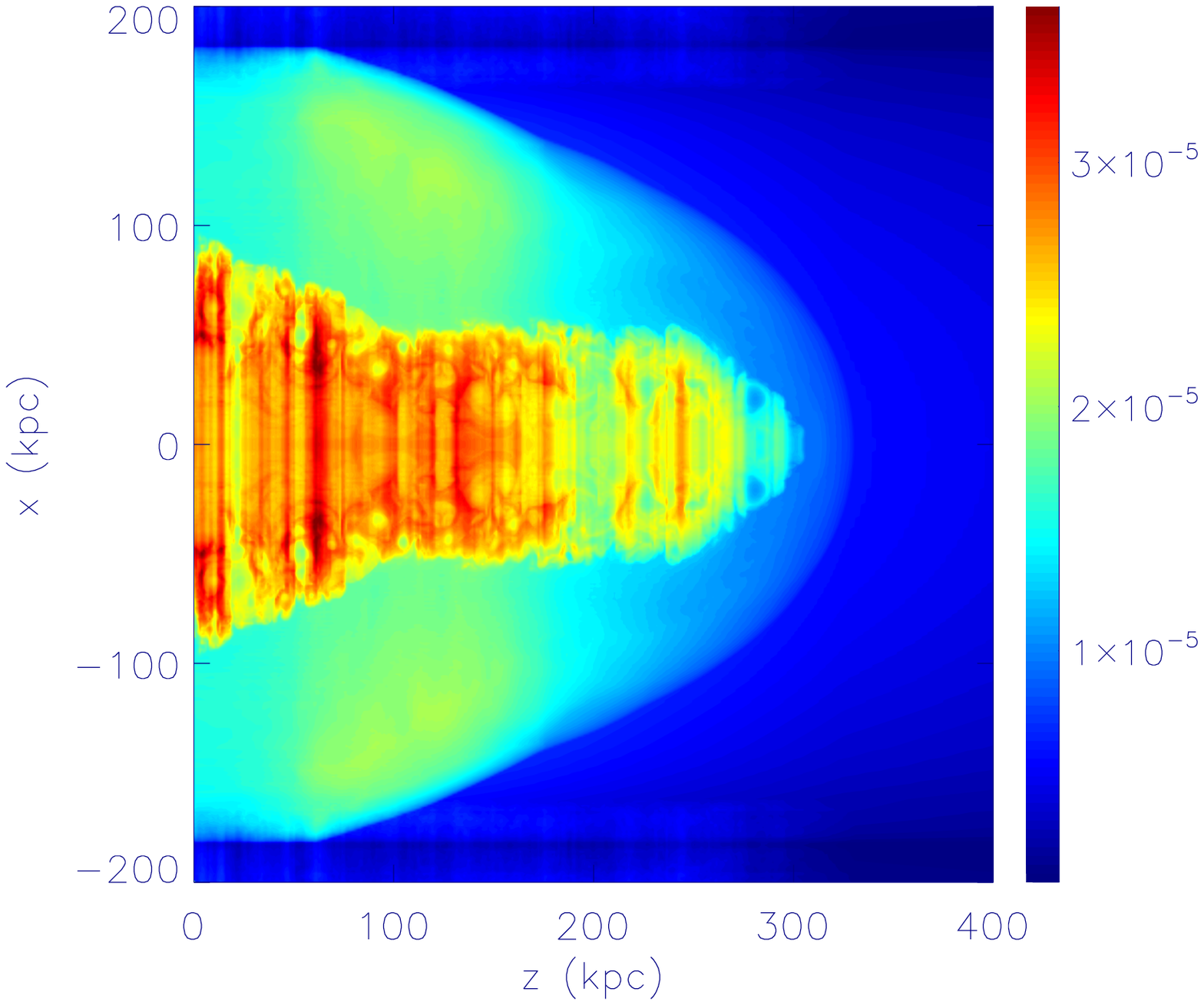}
\includegraphics[angle=0, width=8.5cm]{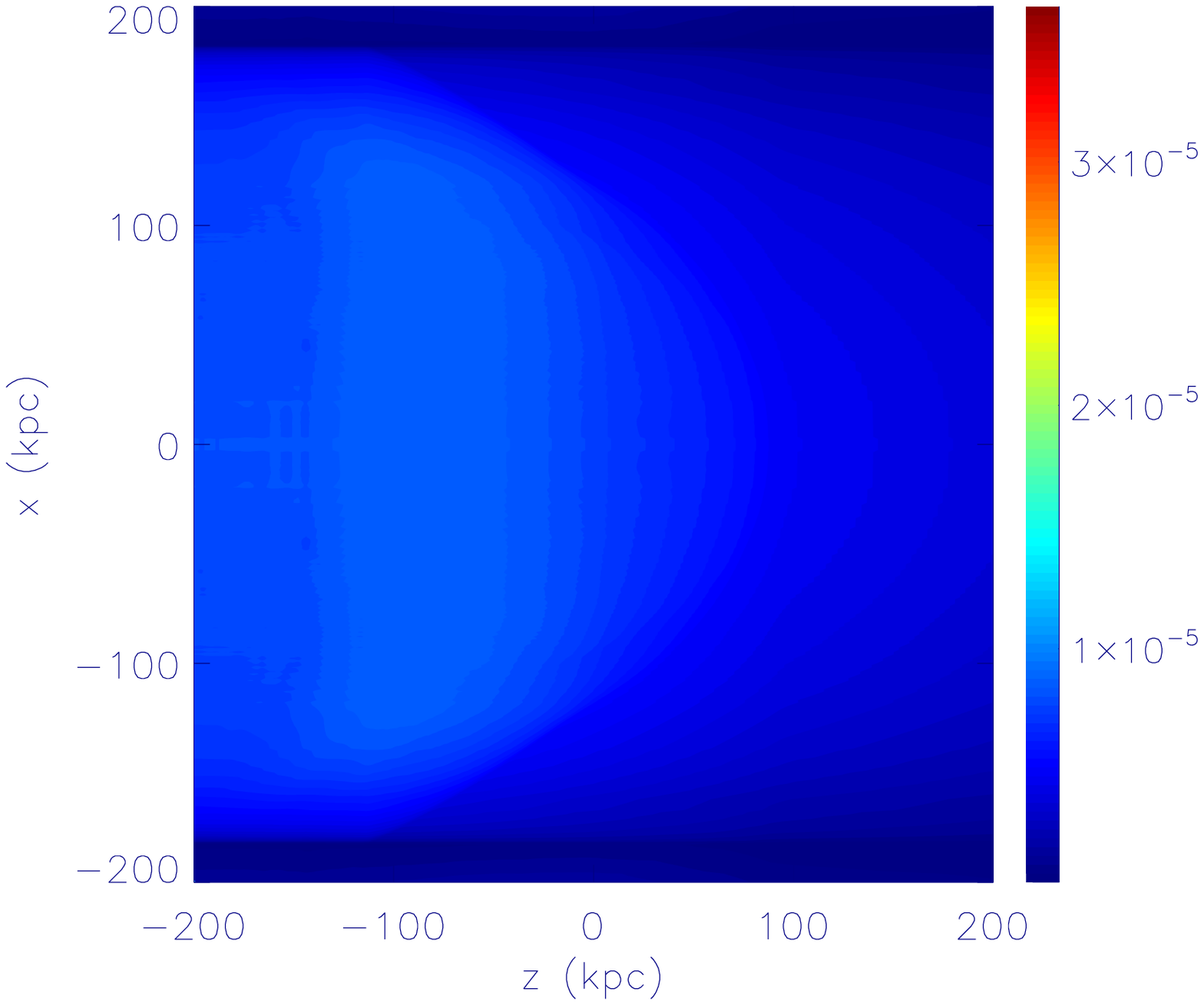}
\includegraphics[angle=0, width=8.5cm]{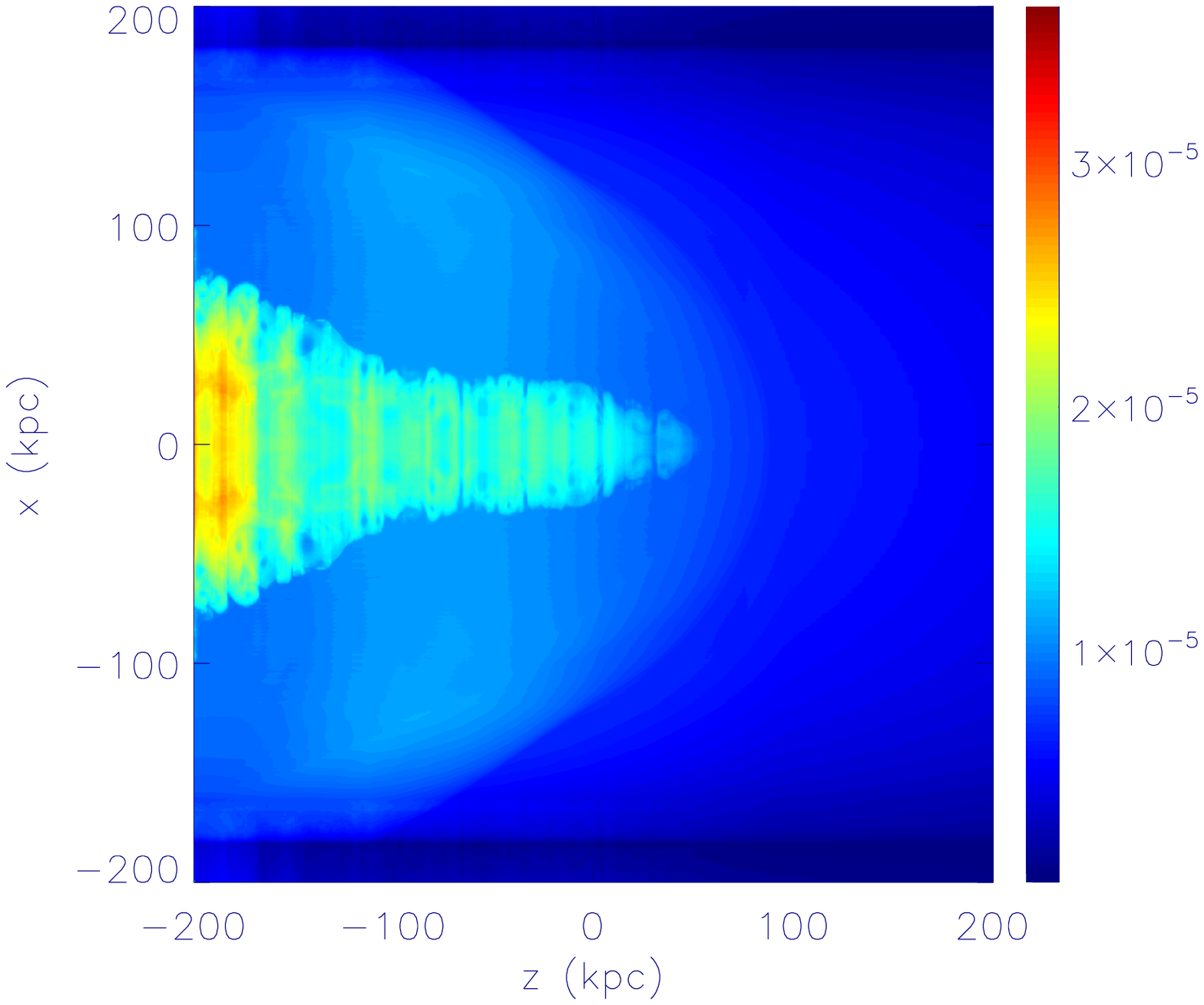}
\caption{The simulated SZ intensity maps, $\Delta I/I_{0}$, at 857 GHz for a viewing angle of 90$^{\circ}$ calculated  
in the frameworks of the Kompaneets approximation (left-hand panels) and the Wright formalism (right-hand panels).
Upper panels correspond to the Model A and bottom panels correspond to the Model B.
}
\label{szp}
\end{figure*}

To describe the Comptonization of the CMB by relativistic electrons, we use the relativistically 
correct formalism proposed by Wright (1979) that is valid for both non-relativistic and relativistic electrons.  
The CMB spectral distortion in the Wright formalism can be written in the form used by Prokhorov et al. (2010) and given by
\begin{equation}
\Delta I(x)=I_{0} \frac{\sigma_{\mathrm{T}}}{m_{\mathrm{e}}c^2} \int n_{\mathrm{e}} k_{\mathrm{b}} T_{\mathrm{e}} G(x, T_{\mathrm{e}}) dl,
\label{eq3}
\end{equation}
with the relativistic spectral function, $G(x, T_{\mathrm{e}})$, as
\begin{equation}
G(x, T_{\mathrm{e}})=\int^{\infty}_{-\infty} \frac{P_{1}(s, T_{\mathrm{e}})}{\Theta(T_{\mathrm{e}})} \left(F(x\exp(-s))-F(x)\right) ds
\end{equation}
where $\Theta(T_{\mathrm{e}})=k_{\mathrm{b}}T_{\mathrm{e}}/m_{\mathrm{e}}c^2$, $F(x)=x^3/(\exp(x)-1)$ 
is the Planck spectrum, and $P_{1}(s, T_{\mathrm{e}})$ is the distribution of frequency shifts for single 
scattering (see Wright 1979).
Note that the relativistic spectral function, $G(x, T_{\mathrm{e}})$, depends on electron temperature and, 
therefore, the spectra of CMB distortions caused by very high temperature gas with  $k_{\mathrm{b}} T_{\mathrm{e}}\simeq 100$ keV 
are significantly different from those obtained in the framework of the Kompaneets approximation. 
The very high temperature gas in AGN cocoons should cause the strong CMB distortions at a frequency of 217 GHz 
and at very high frequencies. The dependence of the relativistic spectral function, $G(x, T_{\mathrm{e}})$ at 
a frequency of 857 GHz on electron temperature is shown in Fig. \ref{F1}. The spectral function at 217 GHz
is shown in fig. 2 of Prokhorov et al. (2010).

\begin{figure*}
\centering
\includegraphics[angle=0, width=8.0cm, height=7.0cm]{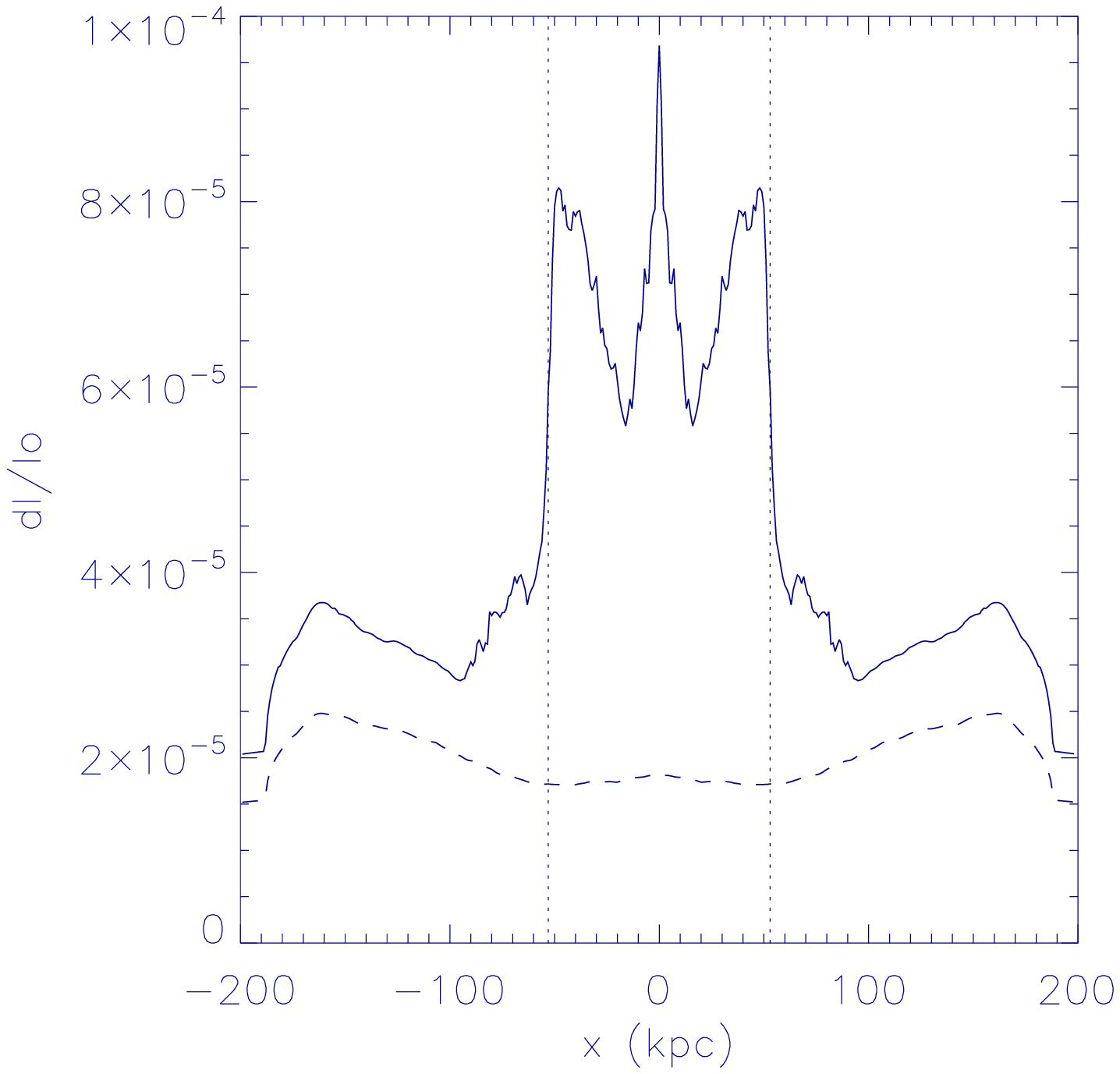}
\includegraphics[angle=0, width=8.0cm, height=7.0cm]{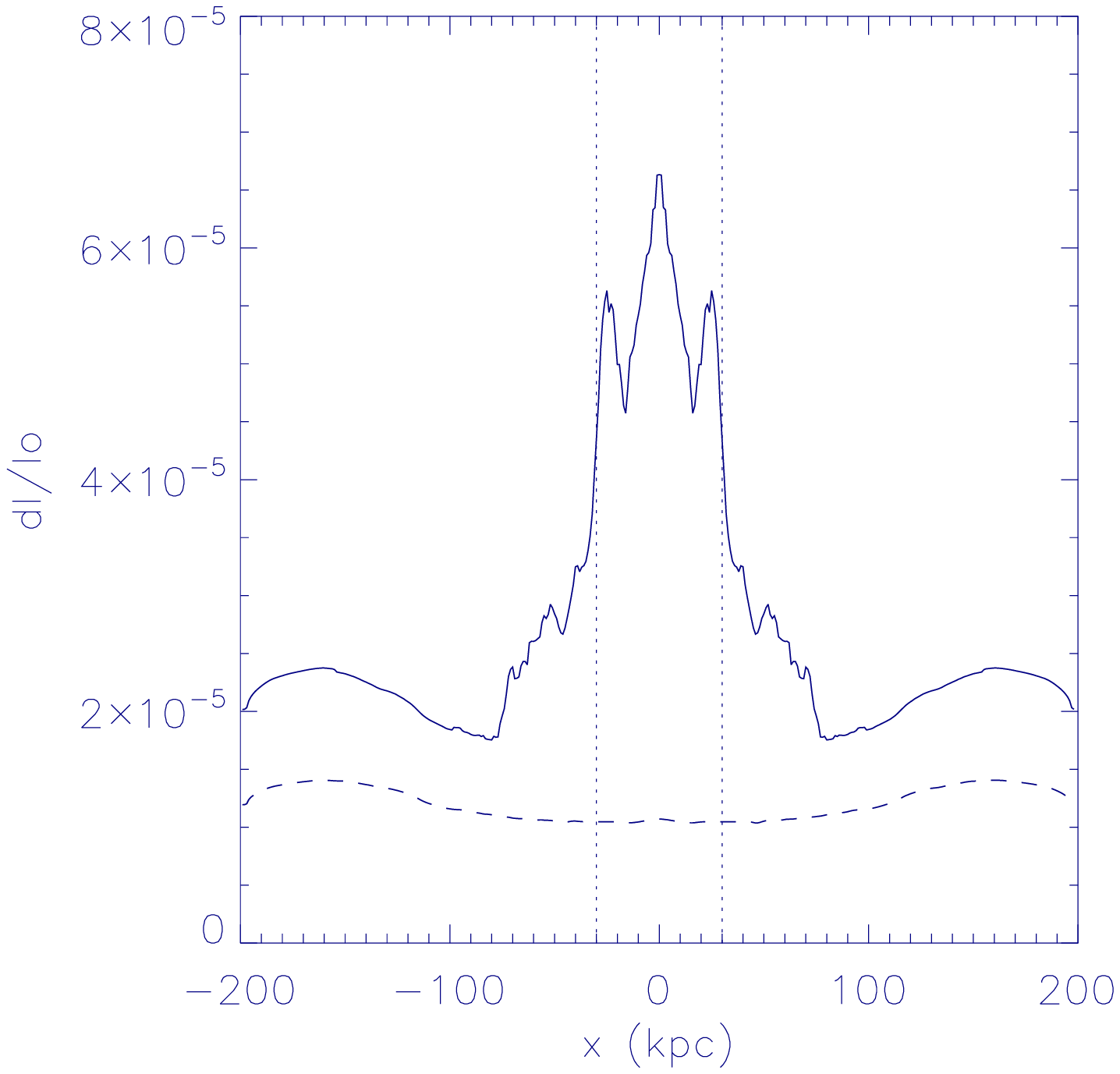}
\caption{The simulated SZ intensity radial profiles, $\Delta I/I_{0}$, at 857 GHz for a viewing angle of 0$^{\circ}$ calculated  
in the frameworks of the Wright and Kompaneets formalisms are shown by solid and dashed curves, respectively.
Left-hand panels correspond to the Model A and right-hand panels correspond to the Model B.
The dotted lines show the boundary surfaces of AGN cocoons. 
} \label{sza}
\end{figure*}

The Kompaneets equation describes a diffusion process in the momentum space and this 
is correct while the energy transferred from an energetic electron to a photon is significantly 
smaller than the initial photon energy. The spectral shape of the CMB distortion in the Kompaneets 
approximation does not depend on the electron temperature and the spectral function, g(x), is null 
at the crossover frequency of 217 GHz ($x=3.83$) and rapidly decreases with frequency at high frequencies ($\nu>600$ GHz).
However, taking into account the relativistic corrections to the SZ effect shows that (1) the relativistically
correct spectral function does not equal zero at a frequency of 217 GHz and that (2) the contribution of high temperature
electrons is significant at higher frequencies. These two properties of the relativistic spectral functions 
are of  interest for finding the hitherto undetected, dynamically-dominant component in the cocoons.
The first property of the relativistic spectral function was used by Colafrancesco (2005) and Pfrommer et al. (2005) 
to disentangle the SZ signal from the X-ray cavities in MS0735+7421 caused by highly energetic electrons from the ambient ICM.  
Below, we use the second property of the relativistic spectral function and show that the SZ signal at high frequencies
from AGN cocoons, owing to the contribution of high temperature electrons, is higher than that from 
the shocked ambient ICM.  

{Note that the relativistically correct Wright formalism (as is the non-relativistic Kompaneets approximation) 
is based on two assumptions: (1) the Thomson cross-section is applicable; and (2) the photons are sufficiently
soft that the energy of the scattered photons is less than that of the electrons. These assumptions are valid for 
calculating the SZ effect produced by high temperature electrons in AGN cocoons (see Loeb 1991; Prokhorov et al. 2010).
The equivalence of different relativistically correct formalisms for calculating the SZ effect under these assumptions 
has been confirmed by Nozawa \& Kohyama (2009). 
Therefore, the results obtained from the Wright formalism are equivalent
to those that can be derived by using other relativistic formalisms for the SZ effect.}

\subsection{SZ intensity maps at high frequencies}

Taking into account the relativistic corrections to the SZ effect is necessary for calculating the SZ effect on AGN cocoons 
produced by electrons with high temperatures ($k_{\rm{b}}T_{\rm{e}}\simeq 10^{9}-10^{10}$ K) such as those derived from
hydrodynamic simulations. 
To produce the 3D pressure and temperature maps which are necessary to derive the intensity map of the SZ effect we rotate
the 2D pressure and temperature maps shown in Figs. \ref{PT1} and \ref{PT2} (for Models A and B, 
respectively) around the jet axis. We calculated the SZ effect using the values of the relativistic spectral
functions found in Sect. 3.1. Since the gas temperatures in a cocoon are high, giant AGN cocoons 
should be a stronger source of the SZ effect at a high frequency (e.g. at $\nu$ = 857 GHz at which the SZ
effect is observable by {\it{Herschel-SPIRE}}) compared with that from the shocked ambient gas. 
Thus, we compare the SZ intensity maps calculated by using the non-relativistic Kompaneets approximation
with those calculated by using the relativistically correct Wright formalism to demonstrate
importance of the consideration of relativistic effects.

We calculated the SZ intensity maps, $\Delta I/I_{0}$, for different viewing angles 
{(note that the intensity of $I_{0}=2(k_{\mathrm{b}} T_{\mathrm{cmb}})^3/(hc)^2$ equals 6.0 mJy arcsec$^{-2}$ and that the SZ intensity does not depend on red-shift of galaxy clusters)}. The SZ intensity maps for
a viewing angle of 90$^{\circ}$ (when the jet is seen from the side) are plotted in Fig. \ref{szp} 
and for a viewing angle of 0$^{\circ}$ (when the jet is seen  face-on) are plotted 
in Fig. \ref{sza}. Comparing the SZ intensity maps calculated in the frameworks of the Kompaneets and 
Wright formalisms, we find a significant difference in the SZ signals from the simulated AGN cocoons.
This difference is explained by the higher values of the relativistic spectral function 
at temperatures of $k_{\rm{b}} T_{\rm{e}}\sim 100$ keV (i.e. $T_{\rm{e}}\sim 10^{9}$ K) compared
with the values of the non-relativistic spectral function. Future observations of the predicted excess of 
the SZ signal at high frequencies from the AGN cocoons owing to the presence of high 
gas temperatures, $T_{\rm{e}}\simeq 10^{9}-10^{10}$ K, are a promising approach to probing the electron populations in AGN cocoons.

Comparing the SZ intensity maps, shown in Figs. \ref{szp} and \ref{sza}, calculated in Model A with those 
calculated in the Model B, we find the presence of high temperature gas in both the simulated AGN cocoons
causes a similar excess of the SZ effect at high frequencies. 

\begin{figure*}
\centering
\includegraphics[angle=0, width=8.5cm]{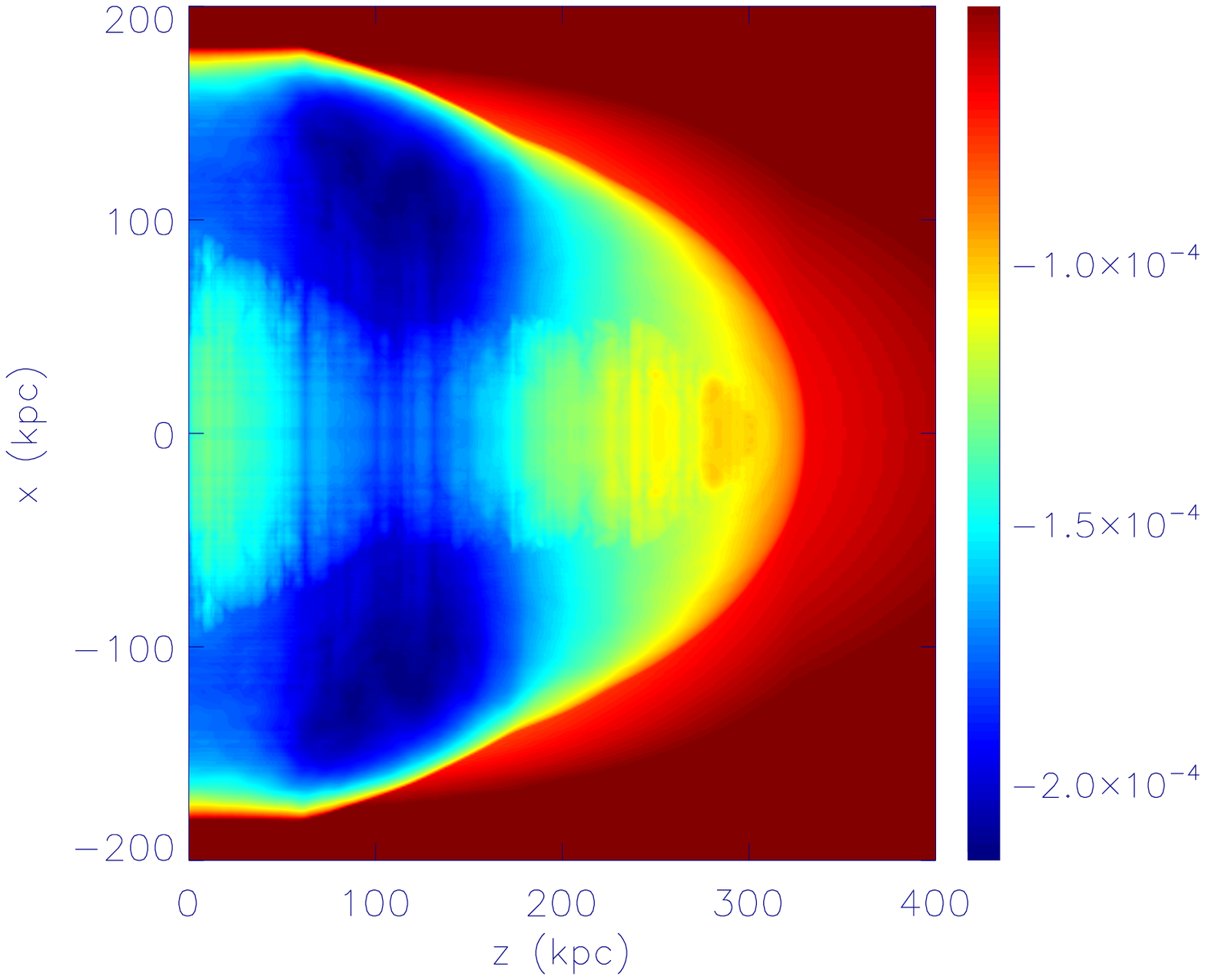}
\includegraphics[angle=0, width=8.5cm]{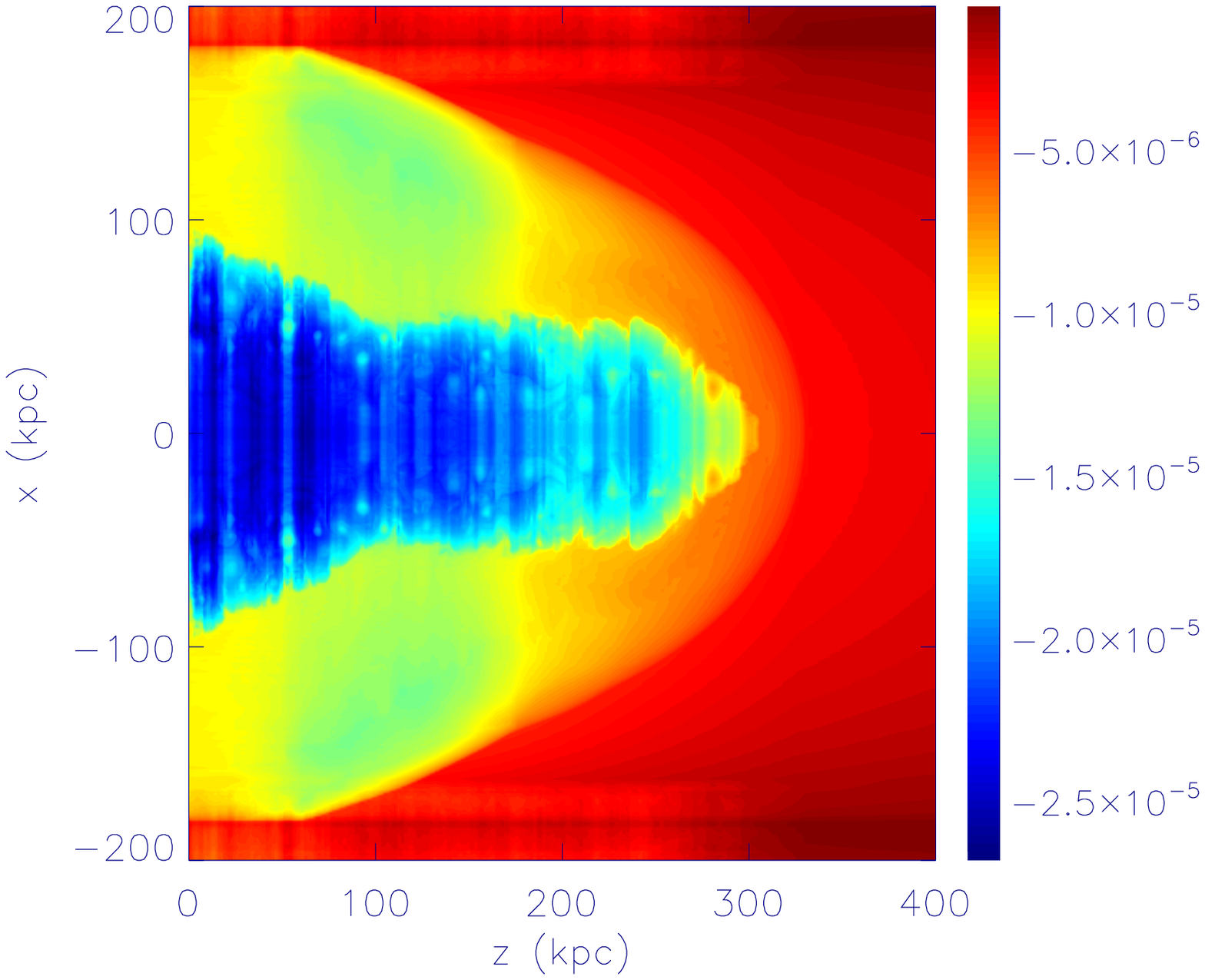}
\caption{The simulated SZ intensity maps, $\Delta I/I_{0}$, for the Model A at 90 GHz (left-hand panel) and at 217 GHz (right-hand panel) 
for a viewing angle of 90$^{\circ}$ calculated in the framework of the Wright formalism 
} \label{sz90}
\end{figure*}

\subsection{SZ intensity maps at 90 and 217 GHz}

In this section, we calculate SZ intensity maps of simulated AGN cocoons at lower frequencies.
For this analysis, we perform SZ map calculations at two frequencies, 90 GHz and 217 GHz, 
at which the possible observations of AGN cocoons using the {\it{GBT}} and {\it{ALMA}} Telescopes 
have been discussed  by Pfrommer et al. (2005), Prokhorov et al. (2010), and Scaife et al. (2010). 

The dependences of spectral functions at frequencies, 90 and 217 GHz, on the temperature 
were calculated in the framework of the Wright formalism for the broad interval of temperatures
in the paper by Prokhorov et al. (2010), see their figs. 1 and 2.   
The absolute value of the spectral function, G(x, T$_{\mathrm{e}}$), at a frequency 90 GHz significantly 
(and monotonically) decreases with electron temperature and a strong deviation from the spectral value derived
in the Kompaneets approximation is expected for gas with $k_{\mathrm{b}} T_{\mathrm{e}}\simeq 100$ keV.
The simple theoretical argument of why the strong deviation of the spectral function 
(at $k_{\mathrm{b}} T_{\mathrm{e}}\gtrsim 100$ keV) derived in the relativistically
correct formalism from that derived in the Kompaneets approximation was presented by Prokhorov et al. (2010).
The frequency of 217 GHz corresponds to the crossover frequency of the non-relativistic SZ effect 
(i.e. derived in the framework of the Kompaneets formalism) and, therefore, the SZ effect from
an AGN cocoon at this frequency is dominated by the high temperature electron component with 
$k_{\mathrm{b}} T_{\mathrm{e}}\gtrsim 10$ keV. Note that the spectral function at the frequency 217 GHz
has a peak at a temperature $k_{\mathrm{b}} T_{\mathrm{e}}= 160$ keV (see Prokhorov et al. 2010).

The intensity maps of the SZ effect at frequencies of 90 GHz and 217 GHz derived from the simulation maps of the gas pressure and temperature in Model A and for a viewing angle 90$^{\circ}$ are plotted in Fig. \ref{sz90}. A comparison of the left-hand and right-hand panels of Fig. \ref{sz90}, which show the SZ intensity at frequencies of 90 GHz and 217 GHz, respectively, demonstrates that the morphology
of the SZ intensity maps at these frequencies are significantly different. The morphology of the SZ map at 217 GHz is similar to that  at 857 GHz. We find that this is the result of the similarity of the spectral functions at 217 Ghz and 857 GHz which are peaked at temperatures of $k_{\rm{b}} T_{\rm{e}}\simeq 100$ keV. 
As for the SZ map morphology at 90 GHz, the AGN cocoon is not clearly seen on the simulated SZ map, because the
brightnesses of the inner regions of the AGN cocoon are significantly different. 
Therefore, we conclude that the SZ observations at 217 GHz and at high frequencies (such as 857 GHz) are 
potentially more suitable for unveiling the presence of AGN cocoons in the ICM if the viewing angle of 
a jet is $\simeq$ 90$^{\circ}$, as of those in the MS 0735+7421 and Hercules A galaxy clusters.
Moreover, an SZ intensity cavity at 90 GHz (see Pfrommer et al. 2005) produced by mildly relativistic electrons 
in an AGN cocoon due to relativistic SZ effect corrections can be reproduced by the presence 
of cosmic ray protons and/or electrons of very high energies which contribute negligibly to the total SZ effect 
(see, e.g., En{\ss}lin \& Kaiser 2000). Since the spatial feature on the SZ maps at 217 GHz and 857 GHz produced
by the presence of an AGN cocoon can be interpreted only as the presence of high temperature gas 
with $k_{\mathrm{b}} T_{\mathrm{e}}\simeq 100$ keV, SZ observations at these frequencies should provide us with
a `smoking gun' for the presence of such high temperature gas in AGN cocoons.

\subsection{{Observational considerations}}

{In this section, we briefly discuss the possibility to observe the SZ effect from AGN cocoons by means of modern instruments} 
({\it Planck-HFI, Herschel-SPIRE,} and {\it ALMA}).
We also show that hard X-ray observations of AGN cocoons can be a test for the presence of nonthermal ultra-relativistic electrons rather than a test for the presence  of very high temperature gas and, therefore, SZ observations are a more direct method to check the presence of high temperature plasma in AGN cocoons. 

\begin{itemize} 
\item \textbf{High-frequency SZ observations}
\end{itemize}

{Our analysis of relativistic spectral functions at high frequencies ($\nu\simeq$857 GHz, 
see Sect. 3.1)  reveals that their global maximuma are at a temperature 
of $k_{\mathrm{b}}T_{\mathrm{e}}\simeq 100$ keV and that these functions have bell shapes.
The  relativistic spectral function at a frequency of 217 GHz also has a bell shape and
peaks at a temperature of $k_{\mathrm{b}}T_{\mathrm{e}}=160$ keV (see Prokhorov
et al. 2010). In this section, we discuss the detectability of AGN cocoons at these
frequencies with modern instrumentation and how to obtain the constraints on thermal energy
stored in AGN cocoons while taking into account the bell shapes of these relativistic spectral functions.} 

{The fact that the absolute values of the relativistic spectral functions at 217 and 857 GHz have global maximuma 
(equal $\approx 1.2$ and $\approx 1.8$, respectively) allows us to derive the lower limit on the pressure, 
$P_{\mathrm{e}}$, integrated along the line-of-sight that is written as (see also} Eq. \ref{eq3})
\begin{equation}
  \int P_{\mathrm{e}} dl\geq \frac{m_{\mathrm{e}} c^2}{\sigma_{\mathrm{T}} G_{\mathrm{max}}
  (x, T_{\mathrm{e}})}\left(\frac{\Delta I}{I_{0}}\right)_{\mathrm{obs}},
  \label{llp}
\end{equation}
{where $G_{\mathrm{max}}$ is the absolute maximal value of a relativistic spectral function
and $\left(\Delta I/{I_{0}}\right)_{\mathrm{obs}}$ is the observed SZ intensity.}

{On the other hand, the upper limit on the contribution of very high temperature electrons
to the pressure integrated along the line-of-sight can be derived from the observed SZ intensity.
Assuming that the electrons contributing to the pressure have temperatures in the range of 
($T_{\mathrm{min}}, T_{\mathrm{max}}$), one can calculate the local minimum of the absolute values of the relativistic spectral functions in the considered temperature range and use the derived local minimum
value to put the upper limit on the contribution of these high temperature electrons to the pressure. The upper limit is given by} 
\begin{equation}
  \int P^{(T_{\mathrm{min}}, T_{\mathrm{max}})}_{\mathrm{e}} dl\leq \frac{m_{\mathrm{e}} c^2}{\sigma_{\mathrm{T}} G_{\mathrm{loc. min.}}
  (x, T_{\mathrm{e}})}\left(\frac{\Delta I}{I_{0}}\right)_{\mathrm{obs/ul}},
  \label{ulp}
\end{equation}
{where the superscript of $(T_{\mathrm{min}}, T_{\mathrm{max}})$ denotes the considered temperature range and $\left(\Delta I/{I_{0}}\right)_{\mathrm{obs/ul}}$ is the observed SZ intensity or upper limit on SZ intensity.
Note that the upper limit on SZ intensity as well as the observed SZ intensity can be used for the purpose of calculating the upper limit on the pressure integrated along the line-of-sight.
Note that the advantage of measuring the SZ effect from AGN cocoons at 217 GHz and 857 GHz 
is that the global absolute maximum of the corresponding relativistic spectral functions coincides with its local absolute maximum in the temperature range of (100 keV, 400 keV).
The local minimum of the relativistic spectral function values at a frequency of 857 GHz for the gas temperatures in the temperature range of $k_{\mathrm{b}}T_{\mathrm{e}}$=(100 keV, 400 keV) equals 0.8 and, therefore, the detection of the SZ signal from an AGN cocoon will provide us with
the lower and upper limits} (see Eqs. \ref{llp} and \ref{ulp}) {on the contribution of 100-400 keV electrons to the pressure that differ by a factor of $G_{\mathrm{max}}/G_{\mathrm{loc. min.}}\approx 2$.}

{The limits on pressure integrated along the line-of-sight can be converted to the limits
on the total thermal energy of high temperature electrons stored in an AGN cocoon. The total thermal energy of electrons, 
$E_{\mathrm{th}}$, is
proportional to the integral, $\int P_{\mathrm{e}} dl$, and can be written as}
\begin{equation}
  E_{\mathrm{th}}=\frac{D^2_{\mathrm{A}}\theta^2}{\gamma-1}\int P_{\mathrm{e}} dl
  \label{eth}
\end{equation}
{where $D_{\mathrm{A}}$ is the angular size distance, $\theta$ is the angular size of a AGN cocoon,
and $\gamma$ is the adiabatic index of the gas. The limits on total thermal energy of electrons can be derived from Eqs.} \ref{llp}, \ref{ulp}, and \ref{eth} {by taking into account the sensitivity of the SZ intruments.}
In Table \ref{tab2}, {we list the upper limits on the thermal energy of high temperature (100-400 keV) electrons that can be derived by means of SZ observations by} {\it Planck-HFI}, {\it Herschel-SPIRE}, and {\it ALMA} {at frequencies of 90 GHz, 217 GHz, and 857 GHz.} {We compared the upper limit listed in this Table with the thermal energy stored in the simulated cocoon (Model A), $E_{\mathrm{th, sim}}\approx7\times10^{60}$ erg, and found that the upper limits on the thermal energy stored in an AGN cocoon that can be obtained with modern SZ observatories are potentially interesting to search for the presence of high temperature gas in AGN cocoons.} 
{Below we describe how to calculate these constraints.}

The {\it Planck} {spacecraft was launched on 14 May 2009 and has been surveying the sky until 15 January 2012 
(the end of its cryogenic lifetime, see Ade et al. 2011). To calculate the limit on the total energy of high temperature electrons that can be obtained by} {\it Planck-HFI}, {we use the sensitivities in units of $\Delta I/I_{0}$ taken from Table 2 of En{\ss}lin \& Kaiser (2000) that were derived using the sensitivities listed in the} {\it Planck-HFI} {proposal. Since the} {\it Planck} {performance parameters determined from flight data (see Ade et al. 2011) do not differ significantly from those that were used by En{\ss}lin \& Kaiser (2000), we do not re-calculate its sensitivity to estimate detectability of the SZ effect from AGN cocoons. }
\addtocounter{footnote}{1}
\footnotetext[\value{footnote}]{where the superscripts of ``${*}$'' and ``${+}$'' denote survey and on-source times, respectively.}
\begin{table*}
\begin{center}
\caption{Upper limits on the thermal energy stored in an AGN cocoon that can be obtained by means of modern instruments}
\label{tab2}
\begin{tabular}{@{}l c c c c c c}
\hline
Instrument & Frequency (GHz) & Angular size ($^{\prime}$) & Integration time$^{\decimal{footnote}}$ & Sensitivity ($\Delta I/I_{0}$) &  Red-shift & $E_{\mathrm{th}, \mathrm{UL}}$ (erg)\\
\hline
{\it Planck-HFI} & 217 & 5.0 & 14 months$^{*}$ & $2.1\times10^{-5}$ &  0.05 & $4.3\times10^{61}$\\
{\it Planck-HFI} & 217 &  5.0 & 14 months$^{*}$ & $2.1\times10^{-5}$  & 0.2 & $4.8\times10^{62}$\\
{\it Planck-HFI} & 857 &  5.0 & 14 months$^{*}$ & $9.7\times10^{-5}$ &  0.05 & $1.6\times10^{62}$\\
{\it Planck-HFI} & 857 & 5.0 & 14 months$^{*}$ & $9.7\times10^{-5}$ &  0.2 & $1.8\times10^{63}$\\
{\it Herschel-SPIRE} & 857 & 0.4 & 4 hours$^{+}$ & $7.0\times10^{-5}$  & 0.05 & $1.0\times10^{60}$\\
{\it Herschel-SPIRE} & 857 & 0.4 & 4 hours$^{+}$ & $7.0\times10^{-5}$ &  0.2 & $1.1\times10^{61}$\\
{\it ALMA} & 90 & 0.5 & 10 hours$^{+}$& $1.7\times10^{-5}$ &  0.05 & $3.5\times10^{59}$\\
{\it ALMA} & 90 & 0.5 &  10 hours$^{+}$& $1.7\times10^{-5}$ &  0.2 & $3.8\times10^{60}$\\
{\it ALMA} & 217 & 0.5 & 120 hours$^{+}$& $4.0\times10^{-5}$ &  0.05 & $8.2\times10^{59}$\\
{\it ALMA} & 217 & 0.5 & 120 hours$^{+}$& $4.0\times10^{-5}$ &  0.2 & $9.1\times10^{60}$\\
\hline
\label{tab2}
\end{tabular}
\end{center}
\end{table*}
{We choose the angular size of an AGN cocoon of $\approx 5^{\prime}$ (that is a bit larger than the FWHM, see Ade et al. 2011) that corresponds to 300 kpc at the red-shift of z=0.05. The calculated upper limits on the thermal energy that can be derived from the} {\it Planck} {observations at 217 GHz and 857 GHz are higher than} $10^{61}$ erg s$^{-1}$. {Although the derived values are
higher than those that can be obtained by the instruments discussed below, the advantage of the} {\it Planck} {mission is that there is the possibility to search for AGN cocoons by surveying the whole sky.}

The {\it Herschel} {spacecraft was launched on 14 May 2009, along with the} {\it Planck} {spacecraft. The} {\it Herschel-SPIRE} {instrument contains an imaging photometer and an imaging spectrometer (Griffin et al. 2006). The imaging photometer operates in three frequency bands centred on 600 GHz, 857 GHz, and 1200 GHz. The first results of observing clusters of galaxies by} {\it Herschel-SPIRE} {has revealed the contribution of relativistic effects to the SZ signal from the Bullet Cluster at a frequency of 857 GHz (Zemcov et al. 2010). To calculate the upper limit on the thermal energy stored in AGN cocoons that can be obtained from the SZ observations at a frequency of 857 GHz with} {\it Heschel-SPIRE}, {we take the instrumental noise level from the paper by Prokhorov et al. (2011) that was derived by using the  the Herschel Observation Planning Tool\footnote{http://herschel.esac.esa.int/Tools.shtml}. The SZ intensity $\Delta I/I_{0}$ of $1.0\times10^{-4}$, that is close to that which was derived in Sect. 3.2 corresponds to $\approx$ 0.4 mJy beam$^{-1}$. The instrument noise of} {\it Herschel-SPIRE} {(Nguyen et al. 2010) can be reduced with integration time.
 We find that a total on-source integration time of 4 h is sufficient to make an instrumental noise level of 0.25 mJy beam$^{-1}$ for a frequency of 857 GHz.  Foreground contamination is a source of additional noise at 857 GHz. The foreground contamination can be subtracted from the 350 $\mu$m (corresponding to 857 GHz) channel by analysing the 250-$\mu$m and 500 $\mu$m emission. Zemcov et al. (2010) generated a 250-$\mu$m source catalogue and corrected the 350- and 500-$\mu$m emission maps to obtain foreground-free measurements of the SZ effect.
We choose the angular size of an AGN cocoon equal to the FWHM that is of $\approx 0.4^{\prime}$ (i.e. 25$^{\prime\prime}$) that corresponds to 75 kpc at the red-shift of z=0.2. The calculated upper limits on the thermal energy of 100-400 keV electrons that can be derived from the} {\it Herschel-SPIRE} {observations at 857 GHz are of the order of} $10^{60}-10^{61}$ erg.

{Modern ground-based SZ observatories, such as} {\it ALMA}\footnote{www.almaobservatory.org/} {(the Atacama Large Millimeter/submillimeter Array), can also provide us with the constraints on the thermal energy of high temperature electrons stored in AGN cocoons.} {\it ALMA} { will consist of a giant array of 12-m antennas and an additional compact array of 7-m and 12-m antennas to greatly enhance ALMA's ability to image extended targets.  High spatial resolution and sensitivity makes this instrument promising for the imaging of AGN cocoons. In the initial phase of} {\it ALMA} { operations, the antennas will be equipped with at least four receiver bands: 84-116 GHz, 211-275 GHz, 275-373 GHz, and 602-720 GHz. Here, we study the possibility to detect the SZ effect from AGN cocoons at frequencies of 90 and 217 GHz with} {\it ALMA}. {We consider only these frequencies because the observations at higher frequencies become excessively time-consuming owing to the narrower beam size and the higher system temperature. To calculate the upper limits on the thermal energy of 100-400 keV electrons that can be obtained with} {\it ALMA}, {we use the ALMA Observation Support Tool\footnote{http://almaost.jb.man.ac.uk/} (Heywood et al. 2011). We found that the sensitivity in units of $\Delta I/I_{0}$ of $1.7\times10^{-5}$  at 90 GHz can be obtained with an integration time of 10 hrs, while the sensitivity of $\Delta I/I_{0}$ of $4.0\times10^{-5}$  at 217 GHz can be obtained with an integration time of 120 hrs. The local minimums of the relativistic spectral functions at 90 GHz and 217 GHz in the temperature range of (100 keV, 400 keV) were taken from Prokhorov et al. (2010) and the angular size of an AGN cocoon was chosen to be equal 0.5$^{\prime}$.
The spatial resolution of} {\it ALMA} {is 5 arcsec at 90 GHz and 2 arcsec at 217 GHz. Therefore,} {\it ALMA} {will allow us to put tighter constraints on the energy of high temperature electrons than those that can be obtained with} {\it Planck} {and} {\it Herschel}, {and to study the internal structure of AGN cocoons}. 

\begin{itemize} 
\item \textbf{Hard X-ray observations}
\end{itemize}

{The SZ effect is not the only mechanism to test the presence of electron populations in AGN cocoons.
Hard X-ray observatories}\footnote{see, http://heasarc.nasa.gov/docs/suzaku/about/overview.html; http://heasarc.gsfc.nasa.gov/docs/astroh/}, {such as {\it Suzaku} and the future mission {\it Astro-H}, are sensitive up to energies of several hundred keV. Here, we compare the hard X-ray luminosities of giant AGN cocoons produced by high temperature gas via bremsstrahlung emission and by nonthermal ultra-relativistic electrons via Inverse Compton (IC) emission and show that the efficiency of IC emission is sigficantly higher than 
that of bremsstrahlung. Therefore, measurements of hard X-rays from AGN cocoons will allow us to test for the presence of nonthermal ultra-relativistic electrons.}

{The luminosity of an AGN cocoon produced by high temperature gas via bremsstrahlung emission is given by}
\begin{equation}
  L_{\mathrm{br}}=\frac{16}{3}\sqrt{\frac{2\pi}{3}}\alpha r_{\mathrm{e}} m_{\mathrm{e}} c^3 n^{2}_{\mathrm{e}} V \sqrt{\frac{k_{\mathrm{b}} T_{\mathrm{e}}}{m_{\mathrm{e}} c^2}} K(T_{\mathrm{e}}),
\end{equation}
{where $\alpha$ is the fine structure constant, $r_{\mathrm{e}}$ is the classical electron radius,
$V$ is the volume of high temperature gas, and $K$ is the corrections for dipole and quadrupole radiation (see, e.g., Khvesyuk et al. 1998) taking into account quantum and relativistic effects. To calculate the value of bremsstrahlung luminosity, we assume that the gas volume equals $10^{71}$ cm$^3$, the gas temperature is $k_{\mathrm{b}}T_{\mathrm{e}}=300$ keV, and the gas density is $10^{-4}$ cm$^{-3}$.
These values are approximately equal to those found in our hydrodynamic simulations.
The calculated value of bremsstrahlung luminosity is $\simeq 2\times10^{41}$ erg s$^{-1}$.}

{To calculate the IC luminosity of the ultra-relativistic electron population that can significantly contribute to the pressure of AGN cocoons, we consider an AGN cocoon filled by ultra-relativistic
electrons with a Lorentz factor of $\tilde\gamma=2\times10^4$ that is required to produce 300 keV photons through
scattering of ultra-relativistic electrons off CMB photons (see, e.g., Ginzburg 1979).
The IC luminosity of the ultra-relativistic electron population is given by (see, e.g., Ginzburg 1979)}
\begin{equation}
L_{\mathrm{IC}}=\frac{4}{3} \frac{\sigma_{\mathrm{T}}\tilde{\gamma}}{m_{\mathrm{e}}c} U_{\mathrm{CMB}} E_{\mathrm{nth.}}
\end{equation}
where $U_{\mathrm{CMB}}$ is the energy density of the CMB, $E_{\mathrm{nth}}$ is the total energy in nonthermal
ultra-relativistic electrons. Assuming the population of nonthermal ultra-relativistic electrons fills the volume of
$10^{71}$ cm$^3$ and its pressure equals that of the simulated AGN cocoon, the calculated value of IC luminosity
is $\simeq 6\times10^{45}$ erg s$^{-1}$ and is four orders of magnitude greater than that of bremsstrahlung emission
from the 300 keV gas. Therefore, it is highly unlikely that the emission of 300 keV gas via bremsstrahlung can be constrained through X-ray 
observations if the dynamically subdominate population of nonthermal ultra-relativistic electrons, that contributes $\gtrsim$0.01\% to the pressure, is present in AGN cocoons.

\section{Conclusions}

Giant X-ray cavities have been discovered by \textit{Chandra} in clusters of galaxies, such as MS 0735+7421 at a redshift z=0.22
(McNamara et al. 2005) and Hercules A at z=0.154 (Nulsen et al. 2005). 
These cavities could be created by powerful AGN outflows with  jet kinetic power of $\gtrsim 10^{46}$ erg s$^{-1}$. 
Similar X-ray surface brightness depression regions have been observed in the galaxy cluster Abell 2204 at a
redshift of 0.152 and the jet's kinetic power of $\simeq 5\times10^{46}$ erg s$^{-1}$ should be necessary 
to create such regions (Sanders et al. 2009). Another example is the intermediate redshift (z=0.29) cluster gas 
associated with the FR II radio galaxy 3C 438 (see Kraft et al. 2007), the observed surface brightness discontinuity 
in the gas that extends $\simeq$600 kpc can be the result of an extremely powerful outburst which is even more powerful 
than those seen in the nearby clusters MS 0735+7421, Hydra A, and Hercules A. 

The standard evolutionary scenario for AGNs suggests that their jets are enveloped in a cocoon (Scheuer 1974;
Blandford \& Rees 1974).
Modern hydrodynamical simulations led to the conclusion that the gas density 
is very low and gas temperature is very high, $10^{9}-10^{10}$ K, in AGN cocoons.  
Therefore, a significant X-ray depression is predicted towards AGN cocoons.
X-ray cavities often coincide with the radio lobes of the central radio galaxy, although
the non-thermal pressure derived from the equipartition condition for the energy of 
synchrotron-radiating non-thermal electrons and magnetic fields is a factor of ten smaller 
than the pressures required to inflate the cavities (e.g. Ito et al. 2008). This
implies that most of the energy in the cocoon is carried by an `invisible' component such as, e.g., 
high energy thermal electrons (Ito et al. 2008). Observations of the SZ effect have been proposed 
to probe the inferred dynamically-dominant component of plasma bubbles associated with X-ray
cavities (see Pfrommer et al. 2005; Colafrancesco 2005; Prokhorov et al. 2010), since
the SZ signal is proportional to the electron pressure integrated along the line-of-sight and to the
value of the spectral function that depends on electron temperature. 

In this paper, we present RHD simulations of the formation and evolution
of AGN cocoons produced by very light powerful jets. Our simulations have been performed {with} 
the publicly available \textsc{PLUTO} computational code. The parameters of the simulations are
listed in Table 1. We have constrained our initial conditions based on the X-ray observations of the powerful AGN outburst in the MS 0735+7421 cluster of galaxies. The goal of our simulations is to provide a realistic model of gas pressure and temperature distributions in AGN cocoons.
The simulated electron pressure and temperature maps are shown in Figs. 1 and 2 for the two
models with different jet lifetimes or `duty cycles'. We have found that the derived temperatures in
AGN cocoons, $\simeq 10^9-10^{10}$ K, are significantly higher than those are 
in the regions of shocked ambient gas and, therefore, the use of
the relativistically correct SZ formalism is necessary to produce the simulated 
SZ maps of AGN cocoons. 

We have presented the first SZ intensity maps of AGN cocoons at various frequencies derived from the relativistic hydrodynamical simulations 
and calculated in the framework of the relativistically correct SZ formalism. This fully relativistic study provides us with more realistic SZ maps of AGN cocoons, more so than those previously obtained by adopting {analytical toy models} of gas pressure and temperature distributions in cocoons. We show that SZ observations at  frequencies of 217 GHz and at 857 GHz will provide us with a method to test the presence of very hot gas in AGN cocoons. This result confirms the previous conclusions based on the {analytical toy models} that SZ maps at 217 GHz could be used to probe the dynamically-dominant component of AGN cocoons. Moreover, our study demonstrates that  two different regions disturbed by AGN activity are present in the simulated domain, namely an AGN cocoon region and a shocked ambient medium region. We show that by taking into account the presence of these two regions leads us to the conclusion that the AGN cocoon is not clearly seen on the simulated SZ map at a frequency of 90 GHz. Therefore, SZ observations at 217 GHz and at higher frequencies, such as 857 GHz, are more suitable for  studying AGN cocoons than those are at lower frequencies. We conclude that fully relativistic simulations of the SZ effect from AGN cocoons are very important, since the spectral properties of the SZ signal should be taken into account in order to produce realistic SZ maps of cocoons.
 
{We have considered the possibility of observing the SZ effect from AGN cocoons by means of modern instruments, such as} {\it Planck-HFI}, {\it Herschel-SPIRE}, and {\it ALMA} {and have derived  upper limits on the total thermal energy of high temperature electrons stored in an AGN cocoon that can be obtained with these instruments} (see Table \ref{tab2}). {We have compared the derived upper limits with the thermal energy stored in the simulated cocoon (Model A), and have found that the upper limits (that can be derived with} {\it Herschel-SPIRE} and {\it ALMA}) {on the thermal energy of 100-400 keV electrons are close to the value of the thermal energy of electrons in the AGN cocoon obtained from our simulations.
Therefore, AGN cocoons are a suitable target for observations with  modern high-resolution SZ instruments.}

%%%%%%%%%%%%%%%%%%%%%%%%%%%%%%%%%%%%%%%%%%%%%%%%%%%%%%%%%%%%%%%%%%%%
\section{Acknowledgements}
%%%%%%%%%%%%%%%%%%%%%%%%%%%%%%%%%%%%%%%%%%%%%%%%%%%%%%%%%%%%%%%%%%%%

Simulations were performed by using a high performance computing cluster 
in the Korea Astronomy and Space Science Institute and 
on the computational cluster belonging to the astronomy department 
at Kyung-Hee University.
AM acknowledges support from the Yonsei University Research Fund 2011 and 2012,
the Korea Astronomy and Space Science Institute Research Fund 2012
and support by the National Research Foundation of Korea to the Center for Galaxy Evolution Research. {We thank the Referee for valuable suggestions.}

\label{lastpage}

\end{document}